\RequirePackage{lineno}
\documentclass[aps,prl,reprint,twocolumn,showpacs]{revtex4-1}

\usepackage{times}
\usepackage{amsmath}
\usepackage{graphicx}
\usepackage{epsfig}
\usepackage{dcolumn}
\usepackage{bm,color}
\usepackage{tabularx}
\usepackage{multirow}
\usepackage{lineno}
\usepackage{longtable}
\usepackage{textcomp}
\usepackage[usenames,dvipsnames,svgnames,table]{xcolor}
\usepackage{appendix}

\newcommand{\BR}{{\cal B}}

\newcommand{\piz}{\pi^0}

\newcommand{\zc}{Z_c(3900)}
\newcommand{\zcp}{Z_c(4020)}
\newcommand{\etac}{\eta_c}

\newcommand{\hc}{h_{c}}
\newcommand{\psip}{\psi(2S)}

\newcommand{\jpsi}{J/\psi}
\newcommand{\ks}{K_{S}^{0}}
\newcommand{\EE}{e^+e^-}
\newcommand{\MM}{\mu^+\mu^-}

\newcommand{\pp}{\pi^+\pi^-}
\newcommand{\kk}{K^+K^-}

\newcommand{\pphc}{\pp h_{c}}
\newcommand{\ppbar}{p\bar{p}}
\newcommand{\pipikk}{\pi^{+}\pi^{-}K^{+}K^{-}}
\newcommand{\pipippbar}{\pi^{+}\pi^{-}p\bar{p}}
\newcommand{\kkkk}{2(K^{+}K^{-})}
\newcommand{\pipipipi}{2(\pi^+\pi^-)}
\newcommand{\pipipipipipi}{3(\pi^+\pi^-)}
\newcommand{\pipipipikk}{2(\pi^+\pi^-)K^{+}K^{-}}
\newcommand{\kskpi}{K^{0}_S K^{\pm}\pi^{\mp}}
\newcommand{\kskpipipi}{K^{0}_{S} K^{\pm}\pi^{\mp}\pi^{+}\pi^{-}}
\newcommand{\kkpiz}{\kk\piz}
\newcommand{\ppbarpiz}{\ppbar\piz}
\newcommand{\kketa}{\kk\eta}
\newcommand{\pipieta}{\pp\eta}
\newcommand{\pipipizpiz}{\pp\piz\piz}
\newcommand{\pipipipieta}{2(\pp)\eta}
\newcommand{\pipipipipizpiz}{2(\pp\piz)}

\newcommand{\ccz}{\chi_{c0}}

\newcommand{\mev}{{\rm MeV}}

\newcommand{\mevcc}{{\rm MeV}/c^2}
\newcommand{\gev}{\mathrm{GeV}}

\newcommand{\gevcc}{\mathrm{GeV}/c^2}

\def\Journal#1#2#3#4{{#1} {\bf #2}, #3 (#4)}

\def\EPJC{Eur. Phys. J. C}

\begin{document}

\title{
\boldmath
Evidence of Two Resonant Structures in $\EE\to\pp\hc$}

\author{
{
M.~Ablikim$^{1}$, M.~N.~Achasov$^{9,e}$, S.~Ahmed$^{14}$,
X.~C.~Ai$^{1}$, O.~Albayrak$^{5}$, M.~Albrecht$^{4}$,
D.~J.~Ambrose$^{44}$, A.~Amoroso$^{49A,49C}$, F.~F.~An$^{1}$,
Q.~An$^{46,a}$
% ...
\\
\vspace{0.2cm}
(BESIII Collaboration)\\
\vspace{0.2cm} {\it
$^{1}$ Institute of High Energy Physics, Beijing 100049, People's Republic of China\\
% ...
$^{4}$ Bochum Ruhr-University, D-44780 Bochum, Germany\\
$^{5}$ Carnegie Mellon University, Pittsburgh, Pennsylvania 15213, USA\\
% ...
$^{9}$ G.I. Budker Institute of Nuclear Physics SB RAS (BINP), Novosibirsk 630090, Russia\\
% ...
$^{14}$ Helmholtz Institute Mainz, Johann-Joachim-Becher-Weg 45, D-55099 Mainz, Germany\\
% ...
$^{44}$ University of Rochester, Rochester, New York 14627, USA\\
% ...
$^{46}$ University of Science and Technology of China, Hefei 230026, People's Republic of China\\
% ...
$^{49}$ (A)University of Turin, I-10125, Turin, Italy; (B)University of Eastern Piedmont, I-15121, Alessandria, Italy; (C)INFN, I-10125, Turin, Italy\\
% ...
$^{a}$ Also at State Key Laboratory of Particle Detection and Electronics, Beijing 100049, Hefei 230026, People's Republic of China\\
% ...
$^{e}$ Also at the Novosibirsk State University, Novosibirsk, 630090, Russia\\
% ...
}
\vspace{0.4cm}
}}

\vspace{0.5cm}

\begin{abstract}

The cross sections of $\EE\to\pphc$ at center-of-mass energies
from 3.896 to 4.600~GeV are measured using data samples
collected with the BESIII detector operating at the Beijing
Electron Positron Collider. The cross sections are found to be of
the same order of magnitude as those of $\EE\to \pp\jpsi$ and
$\EE\to \pp\psip$, but the line shape is inconsistent with the $Y$
states observed in the latter two modes. Two structures 
are observed in the $\EE\to \pp\hc$ cross sections around 4.22 
and 4.39~GeV/$c^{2}$, which we call $Y(4220)$ and $Y(4390)$, 
respectively. A fit with a coherent sum of two Breit-Wigner 
functions results in a mass of $(4218.4^{+5.5}_{-4.5}\pm0.9)$~MeV/$c^{2}$ 
and a width of $(66.0^{+12.3}_{-8.3}\pm0.4)$~MeV for the $Y(4220)$, and a mass of
$(4391.5^{+6.3}_{-6.8}\pm1.0)$~MeV/$c^{2}$ and a width of $(139.5^{+16.2}_{-20.6}\pm0.6)$~MeV 
for the $Y(4390)$, where the first uncertainties are statistical 
and the second ones systematic. The statistical significance of
$Y(4220)$ and $Y(4390)$ is $10\sigma$ over one structure assumption.

\end{abstract}


\maketitle

In the last decade, a series of charmonium-like states have been 
observed at $\EE$ colliders. 
These states challenge
the understanding of charmonium spectroscopy as well as
QCD calculations~\cite{intro-EPJC,INT-review}. According to 
potential models, there are five vector charmonium states 
between the 1D state $\psi(3770)$ and 4.7~$\gevcc$,
namely the $3S$, $2D$, $4S$, $3D$, and $5S$ states~\cite{intro-EPJC}. From
experimental studies, besides the three well-established
structures observed in the inclusive hadronic cross
section~\cite{PDG-2014}, i.e., $\psi(4040)$, $\psi(4160)$,
and $\psi(4415)$, five $Y$ states, i.e., $Y(4008)$, $Y(4230)$, 
$Y(4260)$, $Y(4360)$, and $Y(4660)$ have been reported 
in initial state radiation (ISR) processes $\EE\to\gamma_{\rm
ISR}\pp\jpsi$ or $\EE\to\gamma_{\rm ISR}\pp\psip$ at the
$B$ factories~\cite{intro-BaBar-pipijpsi,intro-Belle-pipijpsi,
intro-BaBar-pipipsip,intro-Belle-pipipsip,intro-BaBar-pipijpsiII,
intro-Belle-pipijpsiII,intro-BaBar-pipipsipII,intro-Belle-pipipsipII} 
or in the direct production processes at the CLEO and BESIII 
experiments~\cite{intro-CLEO-pipijpsi,intro-BESIII-omegachic0}
The overpopulation of structures in this region and the mismatch of
the properties between the potential model prediction and experimental 
measurements make them good candidates for exotic states.
Various scenarios have been proposed, which interpret one or some of 
them as hybrid states, tetraquark states, or molecular 
states~\cite{intro-review-CCLZ}.

The study of charmoniumlike states in different production processes 
supplies useful information on their properties. The process
$\EE\to \pp\hc$ was first studied by CLEO~\cite{intro-CLEO-hcpaper} 
at center-of-mass (c.m.) energies $\sqrt{s}$ from 4.000 to 4.260~GeV. 
A $10\sigma$ signal at 4.170~GeV and a hint of a rising cross
section at 4.260~GeV were observed. Using data samples taken
at 13 c.m. energies from 3.900 to 4.420~GeV~\cite{EX1}, BESIII reported 
the measurement of the cross section of $\EE\to
\pp\hc$~\cite{intro-BESIII-pipihc}. Unlike the line shape of the
process $\EE\to \pp\jpsi$, there is a broad structure in the high
energy region with a possible local maximum at around 4.23~GeV in $\EE\to
\pp\hc$. Based on the CLEO measurement at $\sqrt{s}=4.170$~GeV and
the BESIII measurement, two assumptions were made to
describe the cross section in Ref.~\cite{intro-pipihc-yuancz}. In
both assumptions, a narrow structure exists at around 4.23~GeV,
while the situation in the high energy region is unclear due to the
lack of experimental data.

In this Letter, we present a follow-up study of $\EE\to\pp\hc$ at
c.m. energies from 3.896 to 4.600~GeV using data samples taken at 
79 energy points~\cite{supplemental-material} with the \mbox{BESIII}
detector~\cite{bepc2}. Two resonant structures are observed at 
$\sqrt{s}=4.22$ and 4.39~GeV [hereafter referred to as $Y(4220)$ and 
$Y(4390)$]. The integrated luminosity at each energy 
point is measured with an uncertainty 
of 1.0\% using large-angle Bhabha events~\cite{BESIII-lumi-XYZ2,BESIII-lumi-Rscan}. There are 17
energy points where the integrated luminosities are larger than
40~pb${}^{-1}$ (referred to as ``$XYZ$ data sample" hereafter), while 
the integrated luminosities for the other energy points are smaller than
20~pb${}^{-1}$ (referred to as ``$R$-scan data sample" hereafter). 
The c.m. energies for the $XYZ$ data sample are measured with 
$\EE\to\gamma_{\rm ISR/FSR}\MM$ events with an uncertainty of 
$\pm 0.8$~MeV~\cite{cms-offline-XYZ}, which is dominated by the 
systematic uncertainty. A similar method is used for the $R$-scan 
data sample with multihadron final states~\cite{cms-offline-Rscan}.

In this study, the $\hc$ is reconstructed via its electric-dipole
transition $\hc\to\gamma\etac$ with $\etac\to X_{i}$, where
$X_{i}$ is one of 16 exclusive hadronic final states: $\ppbar$,
$\pipipipi$, $\kkkk$, $\pipikk$, $\pipippbar$, $\pipipipipipi$,
$\pipipipikk$, $\kskpi$, $\kskpipipi$, $\kkpiz$, $\ppbarpiz$,
$\kketa$, $\pipieta$, $\pipipipieta$, $\pipipizpiz$, and
$\pipipipipizpiz$. Here, the $K_S^0$ is reconstructed using its
decay to $\pi^+\pi^-$, and the $\piz$ and $\eta$ from the $\gamma
\gamma$ final state.

Monte Carlo~(MC) simulated events are used to optimize the
selection criteria, determine the detection efficiency, and
estimate the possible backgrounds. The simulation of the BESIII
detector is based on {\sc geant4}~\cite{geant4} and includes the
geometric description of the \hbox{BESIII} detector and the
detector response. For the signal process, we use 
an MC sample for $\EE\to\pp\hc$ process
generated according to phase space.
ISR is simulated with \textsc{kkmc}~\cite{KKMC} with a maximum 
energy for the ISR photon corresponding to the $\pp\hc$ mass
threshold.

\begin{figure}[htbp]
\begin{center}
 \includegraphics[width=0.4\textwidth]{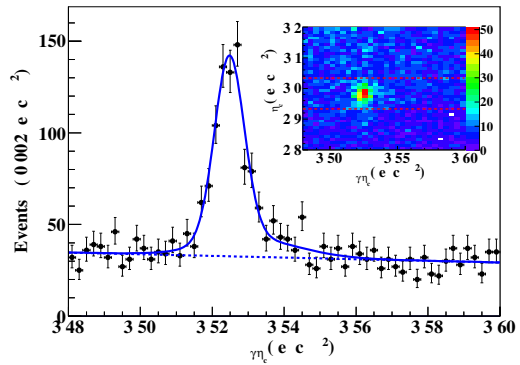}
\caption{ 
The $M_{\gamma\etac}$ distribution in the $\etac$ signal 
region of $4.416~\gev$ data. Points with error bars are the 
data and the curves are the best fit described in the text. The inset
is the scatter plot of the mass of the $\etac$ candidate $M_{\etac}$ 
versus the mass of the $\hc$ candidate $M_{\gamma\etac}$ for the 
same data sample. 
}
\label{fig:4420:mhc}
\end{center}
\end{figure}

We select signal candidates with the same method as that described in
Ref.~\cite{intro-BESIII-pipihc}. Figure~\ref{fig:4420:mhc} shows
the scatter plot of the invariant mass of the $\etac$ candidate
versus the one of the $\hc$ candidate and the invariant mass 
distribution of $\gamma\etac$ in the $\etac$ signal region for 
the data sample at $\sqrt{s}=4.416$~GeV. A clear $\hc\to \gamma\etac$ 
signal is observed. The $\etac$ signal region is defined by a 
mass window around the nominal $\etac$ mass~\cite{PDG-2014}, which
is within $\pm50$~MeV/$c^{2}$ with efficiency about 84\% ($\pm45$~MeV/$c^{2}$ 
with efficiency about 80\%) from MC simulation for final states with 
only charged or $\ks$ particles (for those including $\piz$ 
or $\eta$). 

We determine the number of $\pp\hc$ signal events ($n^{\rm obs}_{\hc}$) 
from the $\gamma\etac$ invariant mass distribution. For the $XYZ$ data
sample, the $\gamma\etac$ mass spectrum is fitted with the MC
simulated signal shape convolved with a Gaussian function to reflect 
the mass resolution difference between the data and MC simulation, 
together with a linear background. The fit to the data sample 
at $\sqrt{s}=4.416$~GeV is shown in Fig.~\ref{fig:4420:mhc}. 
The tail on the high mass side is due to events with ISR (ISR photon undetected); 
this is simulated with \textsc{kkmc} in MC simulation, and its fraction is fixed in the fit. 
For the data samples with large statistics ($\sqrt{s}=4.226$, 4.258,
4.358, and 4.416~GeV), the fit is applied to the 16 $\etac$ decay modes 
simultaneously with the number of signal events in each decay mode constrained 
by the corresponding branching fraction~\cite{bes3-hc-exclusive}. For 
the data samples at the other energy points, we fit the mass spectrum 
summed over all $\etac$ decay modes. 
For the $R$-scan data sample, the number of signal events is calculated 
by counting the entries in the $\hc$ signal region [3.515,~3.535]~GeV$/c^{2}$ 
($n^{\rm sig}$) and the entries in the $\hc$ sideband regions
[3.475,~3.495]~GeV$/c^{2}$ and [3.555,~3.575]~GeV$/c^{2}$ ($n^{\rm
side}$) using the formula $n^{\rm obs}_{\hc}=n^{\rm sig}-fn^{\rm side}$. 
Here, the scale factor $f=0.5$ is the ratio of the size 
of the signal region and the background region, and the background is 
assumed to be distributed linearly in the region of interest.

\begin{figure}[htbp]
\begin{center}
\includegraphics[width=0.4\textwidth]
{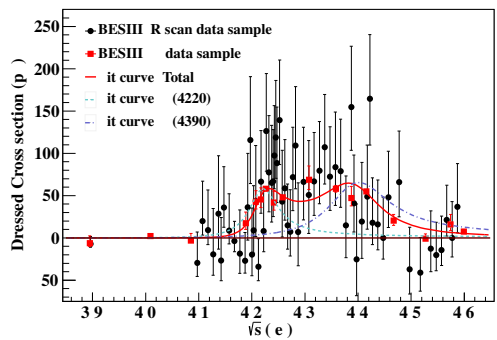}
\end{center}
\caption{Fit to the dressed cross section of $\EE\to\pp\hc$ with
the coherent sum of two Breit-Wigner functions (solid curve). The 
dash (dash-dot) curve shows the contribution from the two structures
$Y(4220)$ [$Y(4390)$].
The dots with error bars are the cross sections for the $R$-scan data sample, 
the squares with error bars are the cross sections for the $XYZ$ data
sample. Here the error bars are statistical uncertainty only.}
\label{cs}
\end{figure}

The Born cross section is calculated from
\begin{equation*}\label{formula-cs}
\sigma^{\rm B}=\frac{n^{\rm obs}_{\hc}}
 {\mathcal{L}(1+\delta)|1+\Pi|^{2}
 \mathcal{B}_{1}
 \sum^{16}_{i=1}\epsilon_{i}\mathcal{B}_{2}(i)},
\end{equation*}

where $n^{\rm obs}_{\hc}$ is the number of observed signal events,
$\mathcal{L}$ is the integrated luminosity, $(1+\delta)$ is the
ISR correction factor, $|1+\Pi|^{2}$ is the correction factor for 
vacuum polarization~\cite{VP-Fred}, 
$\mathcal{B}_{1}$ is the branching fraction of
$\hc\to \gamma\etac$~\cite{PDG-2014},
$\epsilon_{i}$ and $\mathcal{B}_{2}(i)$
are the detection efficiency and branching fraction 
for the $i$ th $\etac$ decay mode~\cite{bes3-hc-exclusive}, respectively.
The ISR correction factor is obtained using the QED calculation 
as described in Ref.~\cite{ISR-factor} and taking the formula used to 
fit the cross section measured in this analysis after two iterations as input.
The Born cross sections are summarized in the Supplemental 
Material~\cite{supplemental-material} together with all numbers 
used in the calculation of the Born cross sections. 
The dressed cross sections (including vacuum polarization effects) are 
shown in Fig.~\ref{cs} with dots and squares for the $R$-scan and $XYZ$ data sample, 
respectively.
The cross sections are of the same order of magnitude as those of the 
$\EE\to\pp\jpsi$ and $\EE\to\pp\psip$~\cite{intro-BaBar-pipijpsi,intro-Belle-pipijpsi,
intro-CLEO-pipijpsi,intro-BaBar-pipipsip,intro-Belle-pipipsip,
intro-BaBar-pipijpsiII,intro-Belle-pipijpsiII,intro-BaBar-pipipsipII,
intro-Belle-pipipsipII}, but follow a different line shape. The
cross section drops in the high energy region, but more slowly than
for the $\EE\to\pp\jpsi$ process.

Systematic uncertainties in the cross section measurement mainly 
come from the luminosity measurement, the branching fraction of $\hc\to
\gamma\etac$, and $\etac\to X_i$, the detection efficiency, the ISR 
correction factor, and the fit. The uncertainty due to the vacuum 
polarization is negligible. The uncertainty in the integrated luminosity 
is 1\% at each energy point. The uncertainty sources for the 
detection efficiency include systematic uncertainties in tracking efficiency (1\% per track), 
photon reconstruction (1\% per photon), and $\ks$ reconstruction (1.2\% per 
$\ks$). Further uncertainties arise from the $\piz$/$\eta$ mass window requirement (1\% per $\piz$/$\eta$), 
the $\chi^{2}_{\rm 4C}$ requirement, $\etac$ parameters, and line shape,
possible intermediate states in the $\pi^{\pm}\hc$ and $\pp$ mass spectra,
intermediate states in $\etac$ decays (included in the uncertainty 
from the branching fraction of $\etac\to X_{i}$), and the limited statistics 
of the MC simulation.

The uncertainty due to the $\chi^{2}_{\rm 4C}$ requirement is estimated by 
correcting the helix parameters of the simulated charged tracks to
match the resolution found in data, and repeating the analysis~\cite{bes3-kinematicfit-eff}.
Uncertainties due to the $\etac$ parameters and line shape are estimated 
by varying them in the MC simulation. When producing 
MC events for the $\EE\to\pp\hc$ process
through the intermediate states $\zc$ or $\zcp$, the parameters of the $\zc$ 
and $\zcp$ are fixed to the average values from the published
measurements~\cite{intro-Belle-pipijpsiII,BESIII-zc3900-pipijpsi,
CLEOc-zc3900-pipijpsi,BESIII-zc3900-DD,intro-BESIII-pipihc}. The
quantum numbers of both $\zc$ and $\zcp$ are assumed to be $J^P=1^{+}$. 
The differences in the efficiency obtained from phase space MC samples 
and those with intermediate $Z_{c}$ states are taken as the uncertainties 
from possible intermediate states in the $\pi^{\pm}\hc$ system. The
uncertainty from intermediate states in the $\pp$ system is estimated 
by reweighting the $\pp$ mass distribution in the phase space MC sample
according to the data, and the resulting difference in the 
efficiency is considered as uncertainty.  The uncertainties due to
data and MC differences in the detection 
efficiency are determined to be between 5.5\% and 10.8\%, depending on the
$\etac$ decay modes and the c.m. energy. Combining the uncertainties 
for the branching fractions of $\etac$ decays~\cite{bes3-hc-exclusive}, the 
uncertainties for the average efficiency 
$\sum_{i=1}^{16}\epsilon_{i}\mathcal{B}_{2}(i)$
are between 6.4\% and 9.1\% depending on the c.m. energy.

The uncertainty in the ISR correction is estimated as described 
in Ref.~\cite{BESIII-zc3900-pipijpsi}. Uncertainties due to the
choice of the signal shape, the background shape, the mass
resolution, and fit range are estimated by changing the $\hc$ and
$\etac$ resonant parameters and line shapes in the MC simulation,
changing the background function from a linear to a second-order
polynomial, changing the mass resolution difference between the data
and the MC simulation by 1 standard deviation, and by extending or
shrinking the fit range. 

Assuming all of the sources are independent, the total systematic 
uncertainty in the $\pphc$ cross section measurement is 
determined to be 9.4\%-13.6\% depending on the c.m. energy. The 
uncertainty in 
$\BR_{1}$ is 11.8\%~\cite{PDG-2014}, 
common to all energy points, and quoted separately in the cross 
section measurement. 
Altogether, the quadratic sum of the common systematic 
errors at each energy point accounts for about 95\% of the total 
systematic error.

A maximum likelihood method is used to fit the dressed cross sections 
to determine the parameters 
of the resonant structures. The likelihood is constructed taking the 
fluctuations of the number of signal and background events into account
(the definition is described in the Supplemental Material~\cite{supplemental-material}).
Assuming that the $\pp\hc$ signal comes from two resonances, the cross
section is parametrized as the coherent sum of two constant width 
relativistic Breit-Wigner functions, i.e.,

\begin{equation*}\label{formula-fit}
\sigma(m)=|B_{1}(m)\cdot\sqrt{\frac{P(m)}{P(M_{1})}}+e^{i\phi}
B_{2}(m)\cdot\sqrt{\frac{P(m)}{P(M_{2})}}|^{2},
\end{equation*}

where 
$B_{j}(m)=\sqrt{12\pi(\Gamma_{ee}\BR)_{j}\Gamma_{j}}/
(m^{2}-M_{j}^{2}+iM_{j}\Gamma_{j})$
with $j=1$ or 2 is the Breit-Wigner 
function, and $P(m)$ is the three-body phase space factor. The masses $M_{j}$, 
the total widths $\Gamma_{j}$, the products of the electronic partial width 
and the branching fraction to $\pp\hc$
($\Gamma_{ee}\BR)_{j}$,
and the relative phase $\phi$ between the two Breit-Wigner functions are
free parameters in the fit. Only the statistical uncertainty is 
considered in the fit. There are two solutions from the fit, one of 
them is shown in Fig.~\ref{cs}. The second solution is very close to 
the one shown here. This can been proved analytically using Eq.(9)
in Ref.~\cite{zhuk-multisolution}, which relates the two solutions 
from the fit when a sum of two coherent Breit-Wigner functions is used.
The parameters determined from the fit are
$M_{1}=(4218.4^{+5.5}_{-4.5})$~MeV$/c^{2}$, $\Gamma_{1}=(66.0^{+12.3}_{-8.3})$~MeV, and
$(\Gamma_{ee}\BR)_{1}= (4.6^{+2.9}_{-1.4})$~eV for $Y(4220)$, 
$M_{2}=(4391.5^{+6.3}_{-6.8})$~MeV$/c^{2}$, $\Gamma_{2}=(139.5^{+16.2}_{-20.6})$~MeV, and
$(\Gamma_{ee}\BR)_{2}=(11.6^{+5.0}_{-4.4})$~eV for $Y(4390)$. 
The relative phase $\phi$ is
$(3.1^{+0.7}_{-0.9})$~rad.  The correlation matrix of the fit parameters
shows large correlation between the $(\Gamma_{ee}\BR)_{j}$
and $\phi$ (see Supplemental Material~\cite{supplemental-material}).

\begin{figure}[htbp]
\begin{center}
\includegraphics[width=0.45\textwidth]
{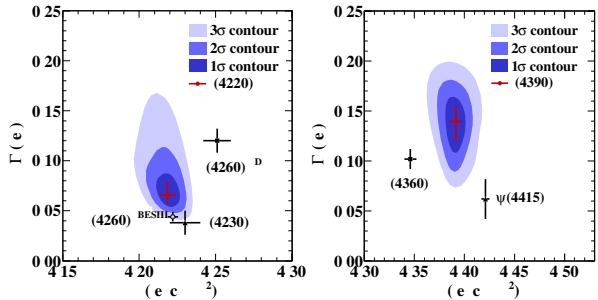}
\end{center}
\caption{
The likelihood contours in the mass and width planes for $Y(4220)$ ({\it left panel})
and $Y(4390)$ ({\it right panel}). The filled areas are up to $3\sigma$ likelihood 
contours and the dots with error bars are the locations of $Y$ or $\psi$ states.
The parameters of $Y(4260)^{\rm PDG}$ are taken from the PDG average~\cite{PDG-2014} 
and $Y(4260)^{\rm BESIII}$ from the measurement of $\EE\to\pp\jpsi$ at 
BESIII~\cite{bes3-pipijpsi}.
}
\label{contour}
\end{figure}

The likelihood contours in the mass and width planes for $Y(4220)$
and $Y(4390)$ are shown in Fig.~\ref{contour}, together with the positions 
of $Y(4230)$, $Y(4260)$, $Y(4360)$, and $\psi(4415)$ with the parameters
taken from the latest PDG average~\cite{PDG-2014}.
The low-lying resonance from the study of $\EE\to\pp\jpsi$ at BESIII~\cite{bes3-pipijpsi}, 
marked as $Y(4260)^{\rm BESIII}$ in the plot, is also compared.  
$Y(4260)$, $Y(4360)$, and $\psi(4415)$ are located outside the $3\sigma$
contours, while $Y(4230)$ and $Y(4260)^{\rm BESIII}$ are overlapped with the 
$3\sigma$ contour of $Y(4220)$.

Fitting the dressed cross section with only one resonance yields 
a worse result, the change of the likelihood value from two resonances to 
one resonance is $[\Delta(-2\mathrm{ln}L)=113.5]$. Taking the change in the 
number of degrees of freedom (4) into account, the significance for
the assumption of two resonant structures over the assumption of one 
resonant structure is $10\sigma$. 
The fit with the coherent sum of one Breit-Wigner function and a 
phase space term gives a worse result as well, the change of the likelihood 
value is $[\Delta(-2\mathrm{ln}L)=66.8]$.
We also fit the cross section with the coherent sum of three Breit-Wigner 
functions, or the coherent sum of two Breit-Wigner functions and a phase space 
term. Both assumptions improve the fit quality, but the significances of 
the third resonance and the phase space term are only $2.6\sigma$ and 
$2.9\sigma$, respectively. 

The systematic uncertainties in the resonance parameters mainly come
from the absolute c.m. energy measurement, the c.m. energy spread, and
the systematic uncertainty on the cross section measurement.  
The uncertainty from the c.m. energy measurement includes
the uncertainty of the c.m. energy and the assumption 
made in the measurement for the $R$-scan data sample.
Because of the low statistics at each energy point in the $R$-scan data
sample, we approximate the difference between the requested and the
actual c.m. energy by a common constant.  To assess the
systematic uncertainty connected with this assumption, we replace the
constant by a c.m. energy-dependent second-order polynomial.
The systematic uncertainty of the c.m. energy is common for all 
the energy points in the two data samples
and will propagate to the mass measurement (0.8~MeV). 
The changes on the parameters are taken as uncertainty. The uncertainty from c.m.
energy spread is estimated by convoluting the fit formula with a
Gaussian function with a width of 1.6~MeV, which is beam spread,
measured by the Beam Energy Measurement System~\cite{BEMS}. 
The uncertainty from the cross section measurement is divided into two
parts. 
The first one is uncorrelated among the different c.m. energy
points and comes mainly from the fit to the $\gamma\etac$ invariant mass 
spectrum to determine the signal yields. The corresponding uncertainty 
is estimated by including the uncertainty in the fit to the cross section, 
and taking the differences on the parameters as uncertainties.
The second part includes all the other sources, is common for all 
data points (14.8\%), and only affects the $\Gamma_{ee}\BR$ measurement.
Table~\ref{tab:sys:Y} summarizes the systematic uncertainty in the
resonance parameters.

\begin{table*}[htbp]
\caption{The systematic uncertainty in the measurement of the
resonance parameters. c.m. energy$_{1,2}$ represent the uncertainty from the 
systematic uncertainty of c.m. energy measurement and the assumption made 
in the c.m. energy measurement for $R$-scan data sample, respectively. Cross 
section$_{a(b)}$ represents the uncertainty from the systematic uncertainties 
of the cross section measurement which are un-correlated (common) in each 
energy point.}\label{tab:sys:Y}
\begin{center}
\begin{tabular}{c|ccc|ccc|c}
\hline\hline
\multirow{2}{*}{Sources}      &    \multicolumn{3}{c|}{$Y(4220)$}    & \multicolumn{3}{c|}{$Y(4390)$} &\multirow{2}{*}{$\phi$ (rad)} \\
                              &~~~$M$ ($\mevcc$)   &~~~$\Gamma$ ($\mev$)& ~~$(\Gamma_{ee}\BR)$ (eV)        
                              &~~~$M$ ($\mevcc$)   &~~~$\Gamma$ ($\mev$)& ~~$(\Gamma_{ee}\BR)$ (eV)        & \\\hline                             
c.m. energy$_{1(2)}$            & $0.8$($0.1$)       & $-$($0.1$)            & $-$($0.2$)
                              & $0.8$($0.1$)       & $-$($0.2$)            & $-$($0.3$)                          & $-$($0.1$)      \\
c.m. energy spread              & $0.1$              & $0.3$                 & $0.3$
                              & $0.1$              & $0.1$                 & $0.7$                               & $0.1$\\
Cross section$_{a(b)}$        & $0.1$($-$)         & $-$($-$)              & $0.2$($0.7$)
                              & $0.6$($-$)         & $0.5$($-$)            & $0.4$($1.7$)                        & $0.1$($-$)\\\hline
Total                         & $0.9$              & $0.4$                 & $0.8$
                              & $1.0$              & $0.6$                 & $1.9$                               & $0.2$ \\
\hline
\end{tabular}
\end{center}
\end{table*}

In summary, we measure the $\EE\to\pp\hc$ Born cross section using data
at 79 c.m. energy points from 3.896 to 4.600~GeV. 
Assuming the $\pp\hc$ events come from
two resonances, we obtain
$M=(4218.4^{+5.5}_{-4.5}\pm0.9)$~MeV/$c^{2}$, 
$\Gamma=(66.0^{+12.3}_{-8.3}\pm0.4)$~MeV, and 
$(\Gamma_{ee}\BR) = (4.6^{+2.9}_{-1.4}\pm0.8)$~eV for $Y(4220)$, 
and $M=(4391.5^{+6.3}_{-6.8}\pm1.0)$~MeV/$c^{2}$, $\Gamma=(139.5^{+16.2}_{-20.6}\pm0.6)$~MeV,
and $(\Gamma_{ee}\BR) =(11.6^{+5.0}_{-4.4}\pm1.9)$~eV for $Y(4390)$, with a
relative phase of $\phi=(3.1^{+0.7}_{-0.9}\pm0.2)$~rad. The first errors
are statistical and the second are systematic. The parameters of these 
structures are different from those of $Y(4260)$, $Y(4360)$, and 
$\psi(4415)$~\cite{PDG-2014}. The resonance parameters of $Y(4220)$ 
are consistent with those of the resonance observed in 
$\EE\to\omega\ccz$~\cite{intro-BESIII-omegachic0}.

The two resonances observed in $\EE\to\pp\hc$ process are located
in the mass region between 4.2 and 4.4~GeV/$c^{2}$, where the vector 
charmonium hybrid states are predicted from various QCD 
calculations~\cite{summary-hybrid-LQCD,summary-hybrid,summary-LQCD-YChen}. 
The mass of $Y(4220)$ is lower than that of $Y(4260)$ observed in the
$\EE\to\pp\jpsi$ process. The smaller mass is consistent with some 
of the theoretical calculations for the mass of $Y(4260)$ when explaining
it as a $D_{1}\bar{D}$ molecule~\cite{summary-molecule-Y4260,summary-lattice-Y4260}.

The BESIII Collaboration thanks the staff of BEPCII and the IHEP
computing center for their strong support. This work is supported in
part by National Key Basic Research Program of China under Contract
No. 2015CB856700; National Natural Science Foundation of China (NSFC)
under Contracts Nos. 11235011, 11322544, 11335008, 11425524; the
Chinese Academy of Sciences (CAS) Large-Scale Scientific Facility
Program; the CAS Center for Excellence in Particle Physics (CCEPP);
the Collaborative Innovation Center for Particles and Interactions
(CICPI); Joint Large-Scale Scientific Facility Funds of the NSFC and
CAS under Contracts Nos. U1232201, U1332201; CAS under Contracts
Nos. KJCX2-YW-N29, KJCX2-YW-N45; 100 Talents Program of CAS; National
1000 Talents Program of China; INPAC and Shanghai Key Laboratory for
Particle Physics and Cosmology; German Research Foundation DFG under
Contracts Nos. Collaborative Research Center CRC 1044, FOR 2359;
Istituto Nazionale di Fisica Nucleare, Italy; Joint Large-Scale
Scientific Facility Funds of the NSFC and CAS	under Contract
No. U1532257; Joint Large-Scale Scientific Facility Funds of the NSFC
and CAS under Contract No. U1532258; Koninklijke Nederlandse Akademie
van Wetenschappen (KNAW) under Contract No. 530-4CDP03; Ministry of
Development of Turkey under Contract No. DPT2006K-120470; NSFC under
Contract No. 11275266; The Swedish Resarch Council; U. S. Department
of Energy under Contracts Nos. DE-FG02-05ER41374, DE-SC-0010504,
DE-SC0012069, DESC0010118; U.S. National Science Foundation;
University of Groningen (RuG) and the Helmholtzzentrum fuer
Schwerionenforschung GmbH (GSI), Darmstadt; WCU Program of National
Research Foundation of Korea under Contract No. R32-2008-000-10155-0.

\onecolumngrid
\appendixpage

\begin{appendix}
\section{Cross section of $\EE\to\pp\hc$}

The number of signal events $n^{\rm obs}_{\hc}$, the luminosity $\mathcal{L}$, 
the product of the initial state radiation correction factor and average efficiency
$(1+\delta)\sum_{i=1}^{16}\epsilon_{i}\mathcal{B}(\etac\to X_{i})$, the vacuum 
polarization correction factor $|1+\Pi|^{2}$, and the Born cross section $\sigma^{\rm B}$ for
$XYZ$ data sample and R-scan data sample are summarized in Table~\ref{XYZ-data}
and Table~\ref{Rscan-data}. The average efficiency is smaller for the R-scan
data sample than for the $XYZ$ data sample due to the different methods used in 
determining the number of signal events.

\begin{table*}[htbp]
\caption{$\EE\to \pphc$ cross sections from $XYZ$ data sample. The
first errors are statistical, and the second ones systematic
uncertainty except the uncertainty in $\BR(\hc\to
\gamma\etac)$, and the third errors are from the
uncertainty in $\BR(\hc\to \gamma\etac)$.} \label{XYZ-data}
\begin{tabular}{crcccccc}
  \hline\hline
  $\sqrt{s}$~(GeV) & ${\cal L}$ (pb$^{-1}$) &~~~~$n^{\rm obs}_{\hc}$~~~~
  & $(1+\delta)\sum_{i=1}^{16}\epsilon_{i}\mathcal{B}(\etac\to X_{i})$(\%) &
  ~~$|1+\Pi|^{2}$~~ & $\sigma^{\rm B}$~(pb) \\
  \hline
  3.8962  &  52.6~~ & $-1.5^{+1.9}_{-0.8}$  & $0.78$ & ~~$1.049$~~ &  $-6.8^{~+8.4}_{~-3.7}~\pm0.6\pm0.8$ \\
  4.0076  & 482.0~~ &~~~$7.6^{+7.2}_{-4.7}$ & $1.42$ & ~~$1.044$~~ &~~~$2.0^{~+2.0}_{~-1.3}~\pm0.3\pm0.2$ \\
  4.0855  &  52.6~~ & $-1.6^{+4.2}_{-1.5}$  & $1.85$ & ~~$1.052$~~ &  $-3.1^{~+8.0}_{~-2.9}~\pm0.4\pm0.4$ \\
  4.1886  &  43.1~~ &~~~$7.4^{+5.7}_{-3.1}$ & $1.96$ & ~~$1.056$~~ &~$16.6^{+12.5}_{~-6.8}\pm1.8\pm2.0$ \\
  4.2077  &  54.6~~ &~$23.6^{+7.2}_{-4.7}$  & $1.99$ & ~~$1.057$~~ &~$40.7^{+12.3}_{~-8.0}\pm3.9\pm4.8$ \\
  4.2171  &  54.1~~ &~$25.3^{+7.8}_{-5.3}$  & $2.03$ & ~~$1.057$~~ &~$43.2^{+13.2}_{~-8.9}\pm4.2\pm5.1$ \\
  4.2263  &1091.7~~ & $669\pm 32$           & $2.07$ & ~~$1.056$~~ & $55.2\pm2.6\pm6.1\pm6.5$ \\ 
  4.2417  &  55.6~~ &~$25.5^{+8.1}_{-5.6}$  & $2.17$ & ~~$1.056$~~ &~$39.9^{+12.5}_{~-8.6}\pm3.8\pm4.7$ \\
  4.2580  & 825.7~~ & $443\pm 27$           & $2.23$ & ~~$1.054$~~ & $46.3\pm2.8\pm5.1\pm5.5$ \\
  4.3079  &  44.9~~ &~$34.3^{+8.3}_{-5.8}$  & $2.22$ & ~~$1.052$~~ &~$65.0^{+15.6}_{-10.9}\pm6.1\pm7.7$ \\
  4.3583  & 539.8~~ & $357\pm 24$           & $2.26$ & ~~$1.051$~~ & $55.8\pm3.7\pm6.2\pm6.6$ \\
  4.3874  &  55.2~~ &~$30.2^{+8.8}_{-6.1}$  & $2.30$ & ~~$1.051$~~ &~$45.5^{+12.9}_{~-9.0}\pm4.4\pm5.4$ \\
  4.4156  &1073.6~~ & $685\pm 35$           & $2.28$ & ~~$1.053$~~ & $52.1\pm2.7\pm5.9\pm6.1$ \\
  4.4671  & 109.9~~ &~$27.1^{+9.7}_{-7.2}$  & $2.38$ & ~~$1.055$~~ &~$19.3^{~+6.9}_{~-5.1}~\pm2.4\pm2.3$  \\
  4.5271  & 110.0~~ & $-1.3^{+7.5}_{-4.9}$  & $2.39$ & ~~$1.055$~~ & $-0.9^{~+5.3}_{~-3.5}~\pm0.1\pm0.1$  \\
  4.5745  &  47.7~~ &~~~$9.2^{+6.8}_{-4.3}$ & $2.37$ & ~~$1.055$~~ &~$15.1^{+11.2}_{~-7.1}~\pm2.1\pm1.8$  \\
  4.5995  & 566.9~~ & $52.0^{+16.9}_{-14.3}$  & $2.38$ & ~~$1.055$~~ &~~~$7.2^{~+2.3}_{~-2.0}~\pm0.7\pm0.8$  \\
  \hline\hline
\end{tabular}
\end{table*}

\begin{longtable*}{cccccccc}
\caption{$\EE\to \pphc$ cross sections in R-scan data sample. The
errors are statistical only. The systematic uncertainty is 18.0\%, and 
common for all energy points.} \label{Rscan-data} \\
 \hline\hline
$\sqrt{s}$~(GeV) & $\mathcal{L}$ (pb${}^{-1}$)  & $n^{\rm sig}$ &
$n^{\rm side}$ & $n^{\rm obs}_{\hc}$ &
$(1+\delta)\sum_{i=1}^{16}\epsilon_{i}\mathcal{B}(\etac\to X_{i})
(\%)$ & $|1+\Pi|^{2}$ & $\sigma^{\rm B}$ (pb) \\ \hline
\endfirsthead

\multicolumn{8}{c}%
{{\bfseries \tablename\ \thetable{} -- continued from previous page}} \\
\hline\hline
$\sqrt{s}$~(GeV) & $\mathcal{L}$ (pb${}^{-1}$)  & $n^{\rm sig}$ & $n^{\rm side}$ & $n^{\rm obs}_{\hc}$ &
$(1+\delta)\sum_{i=1}^{16}\epsilon_{i}\mathcal{B}(\etac\to X_{i})$ (\%) & $|1+\Pi|^{2}$ & $\sigma^{\rm B}$ (pb) \\ \hline
\endhead

\hline \multicolumn{8}{r}{{Continued on next page}} \\ \hline
\endfoot

\hline
\endlastfoot

4.0974 &  7.254 &  0 &  3 & $ -1.5^{+1.9}_{-0.8} $ & 1.37 &  1.052 & $ -28.0^{+34.6}_{-15.3}$\\
4.1074 &  7.146 &  1 &  0 & $~~1.0^{+2.4}_{-0.8} $ & 1.38 &  1.052 & $~~~18.9^{+44.7}_{-15.6}$\\
4.1174 &  7.648 &  1 &  1 & $~~0.5^{+2.6}_{-0.9} $ & 1.39 &  1.052 & $~~~~8.8^{+45.1}_{-16.3}$\\
4.1274 &  7.207 &  1 &  4 & $ -1.0^{+2.8}_{-1.3} $ & 1.40 &  1.053 & $ -18.5^{+51.5}_{-23.4}$\\
4.1374 &  7.268 &  4 &  5 & $~~1.5^{+3.6}_{-2.2} $ & 1.41 &  1.052 & $~~~27.2^{+65.0}_{-39.9}$\\
4.1424 &  7.774 &  1 &  5 & $ -1.5^{+2.9}_{-1.4} $ & 1.42 &  1.052 & $ -25.3^{+48.1}_{-22.9}$\\
4.1474 &  7.662 &  2 &  0 & $~~2.0^{+2.7}_{-1.3} $ & 1.43 &  1.053 & $~~~34.0^{+45.8}_{-21.9}$\\
4.1574 &  7.954 &  1 &  1 & $~~0.5^{+2.6}_{-0.9} $ & 1.45 &  1.054 & $~~~~8.1^{+41.5}_{-14.9}$\\
4.1674 & 18.008 &  2 &  5 & $ -0.5^{+3.1}_{-1.7} $ & 1.46 &  1.054 & $  -3.5^{+22.2}_{-11.9}$\\
4.1774 &  7.309 &  1 &  4 & $ -1.0^{+2.8}_{-1.3} $ & 1.45 &  1.055 & $ -17.5^{+48.8}_{-22.2}$\\
4.1874 &  7.560 &  0 &  3 & $ -1.5^{+1.9}_{-0.8} $ & 1.45 &  1.056 & $ -25.5^{+31.5}_{-13.9}$\\
4.1924 &  7.503 &  4 &  4 & $~~2.0^{+3.5}_{-2.1} $ & 1.45 &  1.057 & $~~~34.2^{+60.3}_{-36.6}$\\
4.1974 &  7.582 &  8 &  3 & $~~6.5^{+4.2}_{-2.9} $ & 1.45 &  1.057 & $ ~109.4^{+70.7}_{-48.6}$\\
4.2004 &  6.815 &  1 &  4 & $ -1.0^{+2.8}_{-1.3} $ & 1.46 &  1.057 & $ -18.7^{+52.0}_{-23.6}$\\
4.2034 &  7.638 &  2 &  3 & $~~0.5^{+3.0}_{-1.5} $ & 1.46 &  1.057 & $~~~~8.3^{+49.9}_{-25.4}$\\
4.2074 &  7.678 &  5 &  5 & $~~2.5^{+3.8}_{-2.4} $ & 1.47 &  1.057 & $~~~41.1^{+62.1}_{-39.7}$\\
4.2124 &  7.768 &  0 &  4 & $ -2.0^{+2.0}_{-1.0} $ & 1.48 &  1.056 & $ -32.4^{+31.6}_{-15.5}$\\
4.2174 &  7.935 &  5 &  2 & $~~4.0^{+3.6}_{-2.3} $ & 1.49 &  1.057 & $~~~62.9^{+57.0}_{-35.4}$\\
4.2224 &  8.212 &  2 &  3 & $~~0.5^{+3.0}_{-1.5} $ & 1.50 &  1.057 & $~~~~7.5^{+45.3}_{-23.0}$\\
4.2274 &  8.193 &  9 &  2 & $~~8.0^{+4.3}_{-3.0} $ & 1.52 &  1.056 & $ ~119.6^{+64.4}_{-45.0}$\\
4.2324 &  8.273 &  8 &  6 & $~~5.0^{+4.3}_{-3.0} $ & 1.53 &  1.056 & $~~~73.4^{+63.5}_{-44.2}$\\
4.2374 &  7.830 &  7 &  6 & $~~4.0^{+4.2}_{-2.8} $ & 1.54 &  1.056 & $~~~61.7^{+64.2}_{-43.8}$\\
4.2404 &  8.571 &  7 &  5 & $~~4.5^{+4.1}_{-2.8} $ & 1.54 &  1.056 & $~~~63.2^{+57.9}_{-39.3}$\\
4.2424 &  8.487 &  7 &  1 & $~~6.5^{+3.9}_{-2.6} $ & 1.54 &  1.056 & $~~~92.1^{+55.7}_{-37.0}$\\
4.2454 &  8.554 & 10 &  4 & $~~8.0^{+4.6}_{-3.3} $ & 1.55 &  1.055 & $ ~112.5^{+63.8}_{-45.7}$\\
4.2474 &  8.596 &  9 &  6 & $~~6.0^{+4.5}_{-3.2} $ & 1.55 &  1.055 & $~~~83.9^{+62.6}_{-44.4}$\\
4.2524 &  8.657 & 12 &  5 & $~~9.5^{+4.9}_{-3.6} $ & 1.55 &  1.054 & $ ~132.0^{+67.4}_{-49.8}$\\
4.2574 &  8.880 &  7 &  8 & $~~3.0^{+4.3}_{-2.9} $ & 1.55 &  1.054 & $~~~40.6^{+57.5}_{-39.7}$\\
4.2624 &  8.629 &  6 &  4 & $~~4.0^{+3.9}_{-2.6} $ & 1.55 &  1.053 & $~~~55.7^{+54.5}_{-35.7}$\\
4.2674 &  8.548 &  3 &  4 & $~~1.0^{+3.3}_{-1.9} $ & 1.55 &  1.053 & $~~~14.0^{+46.5}_{-26.5}$\\
4.2724 &  8.567 &  3 &  5 & $~~0.5^{+3.4}_{-2.0} $ & 1.56 &  1.053 & $~~~~7.0^{+47.0}_{-27.3}$\\
4.2774 &  8.723 &  7 &  4 & $~~5.0^{+4.1}_{-2.8} $ & 1.56 &  1.053 & $~~~68.3^{+55.7}_{-37.6}$\\
4.2824 &  8.596 & 11 &  7 & $~~7.5^{+4.8}_{-3.5} $ & 1.57 &  1.053 & $ ~103.6^{+66.2}_{-48.5}$\\
4.2874 &  9.010 &  6 & 11 & $~~0.5^{+4.2}_{-2.9} $ & 1.57 &  1.053 & $~~~~6.6^{+55.3}_{-38.0}$\\
4.2974 &  8.453 &  8 &  7 & $~~4.5^{+4.4}_{-3.1} $ & 1.57 &  1.052 & $~~~63.0^{+61.1}_{-42.8}$\\
4.3074 &  8.599 &  8 &  9 & $~~3.5^{+4.4}_{-3.1} $ & 1.57 &  1.052 & $~~~48.2^{+61.2}_{-43.2}$\\
4.3174 &  9.342 &  8 &  6 & $~~5.0^{+4.3}_{-3.0} $ & 1.57 &  1.052 & $~~~63.5^{+54.9}_{-38.3}$\\
4.3274 &  8.657 &  7 &  3 & $~~5.5^{+4.0}_{-2.7} $ & 1.57 &  1.051 & $~~~75.5^{+55.4}_{-37.1}$\\
4.3374 &  8.700 &  9 &  3 & $~~7.5^{+4.4}_{-3.1} $ & 1.58 &  1.051 & $ ~102.0^{+59.2}_{-41.6}$\\
4.3474 &  8.542 &  7 &  4 & $~~5.0^{+4.1}_{-2.8} $ & 1.59 &  1.051 & $~~~68.8^{+56.1}_{-37.9}$\\
4.3574 &  8.063 &  8 &  5 & $~~5.5^{+4.3}_{-3.0} $ & 1.60 &  1.051 & $~~~79.5^{+61.9}_{-42.9}$\\
4.3674 &  8.498 &  8 &  5 & $~~5.5^{+4.3}_{-3.0} $ & 1.61 &  1.052 & $~~~74.9^{+58.3}_{-40.5}$\\
4.3774 &  8.158 &  5 &  8 & $~~1.0^{+3.9}_{-2.6} $ & 1.62 &  1.052 & $~~~14.1^{+55.3}_{-36.3}$\\
4.3874 &  7.460 & 10 &  1 & $~~9.5^{+4.4}_{-3.1} $ & 1.61 &  1.051 & $ ~147.2^{+68.3}_{-48.6}$\\
4.3924 &  7.430 &  4 &  3 & $~~2.5^{+3.5}_{-2.1} $ & 1.61 &  1.051 & $~~~39.0^{+54.2}_{-32.4}$\\
4.3974 &  7.178 &  4 & 11 & $ -1.5^{+3.9}_{-2.5} $ & 1.61 &  1.052 & $ -24.2^{+62.2}_{-40.6}$\\
4.4074 &  6.352 &  4 &  6 & $~~1.0^{+3.6}_{-2.3} $ & 1.60 &  1.053 & $~~~18.3^{+66.4}_{-41.3}$\\
4.4174 &  7.519 &  5 &  4 & $~~3.0^{+3.7}_{-2.4} $ & 1.60 &  1.053 & $~~~46.4^{+57.7}_{-36.6}$\\
4.4224 &  7.436 & 11 &  2 & $~10.0^{+4.6}_{-3.3} $ & 1.60 &  1.053 & $ ~156.2^{+71.9}_{-52.0}$\\
4.4274 &  6.788 &  2 &  2 & $~~1.0^{+2.9}_{-1.4} $ & 1.61 &  1.053 & $~~~17.1^{+50.2}_{-24.6}$\\
4.4374 &  7.634 &  3 &  4 & $~~1.0^{+3.3}_{-1.9} $ & 1.62 &  1.055 & $~~~15.0^{+49.8}_{-28.4}$\\
4.4474 &  7.677 &  2 &  4 & $~~0.0^{+3.1}_{-1.6} $ & 1.63 &  1.055 & $~~~~0.0^{+45.6}_{-23.9}$\\
4.4574 &  8.724 &  6 &  5 & $~~3.5^{+4.0}_{-2.6} $ & 1.64 &  1.055 & $~~~45.5^{+51.5}_{-34.0}$\\
4.4774 &  8.167 &  7 &  5 & $~~4.5^{+4.1}_{-2.8} $ & 1.64 &  1.055 & $~~~62.4^{+57.2}_{-38.8}$\\
4.4974 &  7.997 &  2 &  9 & $ -2.5^{+3.3}_{-2.0} $ & 1.65 &  1.055 & $ -35.3^{+47.1}_{-27.7}$\\
4.5174 &  8.674 &  1 &  8 & $ -3.0^{+3.0}_{-1.6} $ & 1.65 &  1.055 & $ -38.9^{+39.2}_{-20.9}$\\
4.5374 &  9.335 &  3 &  8 & $ -1.0^{+3.5}_{-2.1} $ & 1.65 &  1.055 & $ -12.0^{+42.3}_{-25.8}$\\
4.5474 &  8.765 &  1 &  5 & $ -1.5^{+2.9}_{-1.4} $ & 1.65 &  1.054 & $ -19.3^{+36.6}_{-17.5}$\\
4.5574 &  8.259 &  1 &  4 & $ -1.0^{+2.8}_{-1.3} $ & 1.65 &  1.055 & $ -13.7^{+38.1}_{-17.3}$\\
4.5674 &  8.390 &  2 &  1 & $~~1.5^{+2.9}_{-1.4} $ & 1.64 &  1.055 & $~~~20.2^{+38.7}_{-18.3}$\\
4.5774 &  8.545 &  2 &  4 & $~~0.0^{+3.1}_{-1.6} $ & 1.64 &  1.055 & $~~~~0.0^{+40.7}_{-21.3}$\\
4.5874 &  8.162 &  4 &  3 & $~~2.5^{+3.5}_{-2.1} $ & 1.63 &  1.055 & $~~~34.7^{+48.3}_{-28.9}$\\
\end{longtable*}

\newpage

\section{Definition of likelihood function}
In the maximum likelihood fit to the dressed cross sections of $\EE\to\pp\hc$, the likelihood 
is defined as:
\begin{equation}
L(\mu^{\rm sig};p) = \prod_{i=1}^{17}L_{i}(\mu^{\rm sig};p_{i})\prod_{j=1}^{62}L_{j}(\mu^{\rm sig};p_{j}),
\end{equation} 
where $\mu^{\rm sig}$ is the expected number of signal events, $p_{i}$ and $p_{j}$ 
are the parameters in the likelihood functions. $L_{i}$ and $L_{j}$ are the 
likelihood functions for the $XYZ$ and R-scan data samples,
respectively.

The likelihood functions for the $XYZ$ and R-scan data samples are defined
differently due to the different statistics of the samples.  For the
$XYZ$ data sample, 
where the statistics is large, the likelihood function is described by
an asymmetric Gaussian function
$L_{i}=G(\mu^{\rm sig}_{i}; m_{i},\sigma_{i}^{\rm
  left},\sigma_{i}^{\rm right})$,
\begin{eqnarray}
G(\mu_{i}^{\rm sig};m_{i},\sigma_{i}^{\rm left},\sigma_{i}^{\rm right}) & = & \frac{1}{\sqrt{2\pi}
(\sigma_{i}^{\rm left}+\sigma_{i}^{\rm right})}e^{-\frac{(\mu_{i}^{\rm sig}-m_{i})^{2}}
{2(\sigma_{i}^{\rm left})^{2}}}, \mu_{i}^{\rm sig}<m_{i};\nonumber\\
& & \frac{1}{\sqrt{2\pi}
(\sigma_{i}^{\rm left}+\sigma_{i}^{\rm right})}e^{-\frac{(\mu_{i}^{\rm sig}-m_{i})^{2}}
{2(\sigma_{i}^{\rm right})^{2}}}, \mu_{i}^{\rm sig}\geq m_{i};
\end{eqnarray}
By scanning the number of $\hc$ signal events in the fit to the $\gamma\etac$
invariant mass spectrum, the likelihood value as a function of expected signal
events $\mu_{i}^{\rm sig}$ is obtained. The parameters of the Gaussian function 
are determined from a fit to the likelihood distribution. For the R-scan data sample, 
the likelihood is defined as:
\begin{equation}
L_{j}(\mu_{j}^{\rm sig};N_{j}^{\rm sig}, N_{j}^{\rm side}) = \int_{0}^{\infty}P(N_{j}^{\rm sig};
\mu_{j}^{\rm sig}+f\cdot\mu_{j}^{\rm bkg})P(N_{j}^{\rm side};\mu_{j}^{\rm bkg})d\mu_{j}^{\rm bkg},
\end{equation} 
where $P(N;\mu)=\frac{1}{N!}\mu^{N}e^{-\mu}$ is the Poisson distribution, 
$\mu_{j}^{\rm sig}$ and $\mu_{j}^{\rm bkg}$ are the number of expected number 
of signal and background events, respectively.

\section{Correlation matrix from the fit to the cross section}
The correlation matrix of the fit parameters from the fit the to the 
dressed cross sections of $\EE\to\pp\hc$ is
\begin{figure}[ht]
\centering
$\bordermatrix{
      & M_{1} & \Gamma_{1} & (\Gamma_{ee}\mathcal{B})_{1} & \phi & M_{2} & \Gamma_{2} & (\Gamma_{ee}\mathcal{B})_{2} \cr
M_{1} &1.000  &  -0.420    & 0.104                        & 0.136 &~~~0.184  & -0.449     & ~~~0.078 \cr 
\Gamma_{1}  & & ~~~1.000   & 0.191                        & 0.077 &~~~0.478  &~~~0.096    & ~~~0.079 \cr
(\Gamma_{ee}\BR)_{1} & &   & 1.000                        & 0.992 &~~~0.020  & -0.102     & ~~~0.985 \cr
\phi  &       &            &                              & 1.000 &-0.049    & -0.119     & ~~~0.993 \cr
M_{2} &       &            &                              &       & ~~~1.000 & -0.306     & -0.113\cr
\Gamma_{2}  & &            &                              &       &          &~~~1.000    & -0.017\cr
(\Gamma_{ee}\BR)_{2} & &   &                              &       &          &            & ~~~1.000 \cr 
}$
\end{figure}\\
where $M_{j}$, $\Gamma_{j}$, and $(\Gamma_{ee}\BR)_{j}$ ($j=1,~2$) are the mass, the total width, 
and the product of the electronic partial and the branching fraction to $\pp\hc$ for the two 
resonances, respectively; $\phi$ is the relative phase between two Breit-Wigner functions. 
\end{appendix}


\begin{thebibliography}{**}
\bibitem{EX1} The data samples are the same as those used in this
study except at 4.420 GeV. At 4.420 GeV, more data has been collected
and analysed. Besides, the CM energies at all energy points have been
remeasured using $\EE\to\gamma_{\rm ISR/FSR}\MM$ events, and found to
be a few MeV smaller.
\end{thebibliography}
\end{document}